\documentclass[reprint,amsmath,amssymb,aps,pra,showpacs,longbibliography,superscriptaddress]{revtex4-1}
\usepackage[latin9]{inputenc}
\usepackage{mathrsfs}
\usepackage{amsmath}
\usepackage{amssymb}
\usepackage{graphicx}
\usepackage{esint}

\makeatletter

\newcommand*\LyXThinSpace{\,\hspace{0pt}}

\usepackage{epsfig}
\usepackage{bm}
\usepackage{dcolumn}
\usepackage{color}

\makeatother

\begin{document}
\title{Breakdown of the single-mode description of ultradilute quantum droplets
in binary Bose mixtures: A perspective from a microscopic bosonic
pairing theory }
\author{Hui Hu}
\affiliation{Centre for Quantum Technology Theory, Swinburne University of Technology,
Melbourne, Victoria 3122, Australia}
\author{Jia Wang}
\affiliation{Centre for Quantum Technology Theory, Swinburne University of Technology,
Melbourne, Victoria 3122, Australia}
\author{Han Pu}
\affiliation{Department of Physics and Astronomy, Smalley-Curl Institute, Rice
University, Houston, Texas 77251-1892, USA}
\author{Xia-Ji Liu}
\affiliation{Centre for Quantum Technology Theory, Swinburne University of Technology,
Melbourne, Victoria 3122, Australia}
\date{\today}
\begin{abstract}
In his seminal proposal of quantum droplets in binary Bose mixtures
{[}Phys. Rev. Lett. \textbf{115}, 155302 (2015){]}, Dmitry Petrov
suggested that the density ratio $n_{2}/n_{1}$ of the two bosonic
components are locked to an optimal value, which is given by the square
root of the ratio of the two intra-species scattering lengths, i.e.,
$\sqrt{a_{11}/a_{22}}$. Due to such a density locking, quantum droplets
can be efficiently described by using an extended Gross--Pitaevskii
equation within the single-mode approximation. Here, we find that
this single-mode description necessarily breaks down in the deep quantum
droplet regime, when the attractive inter-species scattering length
$a_{12}$ significantly deviates away from the threshold of mean-field
collapse (i.e., $-\sqrt{a_{11}a_{22}}$). By applying a bosonic pairing
theory, we show that the density ratio is allowed to fluctuate in
a sizable interval. Most importantly, the optimal density ratio would
be very different from $\sqrt{a_{11}/a_{22}}$, in the case of unequal
intra-species scattering lengths ($a_{11}\neq a_{22}$). Our finding
might provide a plausible microscopic explanation of the puzzling
low critical particle number of quantum droplets, as experimentally
observed. Our predicted interval of the density ratio, as a function
of the inter-species scattering length, could also be experimentally
examined in cold-atom laboratories in the near future. 
\end{abstract}
\maketitle

\section{Introduction}

Ultradilute quantum droplets are new state of matter recently discovered
in cold-atom experiments \citep{Schmitt2016,Chomaz2016,Cabrera2018,Semeghini2018,Petrov2018,Bottcher2021,Luo2021}.
In comparison to the well-studied helium droplets of nanometer size
at Kelvin temperatures \citep{Harms1998,Barranco2006,Gessner2019},
these self-bound quantum objects have a much larger size (i.e., in
micrometer), and are a million times more dilute and a trillion times
colder \citep{Petrov2018}. Owing to the unprecedented controllability
of cold atoms \citep{Bottcher2021}, ultradilute quantum droplets
serve as a new ideal platform to test theories of quantum many-body
interactions and are shedding new lights to the understanding of quantum
liquids in the strongly interacting regime.

The first ultradilute quantum droplet was observed in a cloud of bosonic
dipolar atoms \citep{Schmitt2016,Chomaz2016}, such as dysprosium
$^{164}$Dy \citep{Schmitt2016}. This unexpected discovery sparked
enormous interests in clarifying the formation mechanism of quantum
droplets \citep{Baillie2016,Wachtler2016}. The most innovative explanation
was given by Dmitry Petrov \citep{Petrov2015}, who proposed the stabilization
of quantum droplets in binary Bose mixtures by the repulsive Lee-Huang-Yang
(LHY) quantum fluctuations \citep{LeeHuangYang1957}, one year before
the observation of dipolar quantum droplets. In 2018, Petrov's pioneering
idea was successfully confirmed by Cabrera \textit{et al.} in Barcelona
\citep{Cabrera2018} and Semeghini \textit{et al.} in Florence \citep{Semeghini2018}.
These two experimental realizations \citep{Cabrera2018,Semeghini2018},
together with the initial observation of dipolar quantum droplets
\citep{Schmitt2016,Chomaz2016} and later developments \citep{Cheiney2018,Ferioli2019,DErrico2019,Cavicchioli2024},
have opened an exciting new research field in quantum many-body physics
and motivated tremendous theoretical investigations of this new state
of matter \citep{Baillie2016,Wachtler2016,Cappellaro2017,Cikojevic2018,Staudinger2018,Ancilotto2018,Parisi2019,Cikojevic2019,Minardi2019,Hu2020a,Tylutki2020,Hu2020b,Hu2020c,Hu2020d,Gu2020,Cikojevic2020,Wang2020,Cikojevic2021,Zin2021,Hu2021,Wang2021,Pan2022,He2023,Yogurt2023,Sanuy2024}.

An important feature of the Petrov theory for quantum droplets in
binary Bose mixtures is the density locking of the two components
\citep{Petrov2015}. Let us denote the two repulsive intra-species
scattering lengths as $a_{11}>0$ and $a_{22}>0$, respectively, and
the attractive inter-species scattering length as $a_{12}<0$. Near
the threshold of mean-field collapse \citep{Petrov2015}, where $a_{12}$
approaches $a_{12}^{(c)}=-\sqrt{a_{11}a_{22}}$ from below, the ratio
between the density $n_{1}$ of the first component and the density
$n_{2}$ of the second component is predicted to be fixed at an optimal
value, i.e., 
\begin{equation}
\frac{n_{2}}{n_{1}}=\sqrt{\frac{a_{11}}{a_{22}}}.\label{eq:DensityRatioPetrov}
\end{equation}
The deviation away from this optimal ratio is energetically prohibited
and only a slight variation $\delta n_{i}/n_{i}\sim\left|\delta a\right|/a_{ii}$
($i=1,2$) is allowed \citep{Petrov2015,Staudinger2018,Ancilotto2018,Zin2021},
where $\delta a\equiv a_{12}-a_{12}^{(c)}$. The prediction of the
density locking relies on the use of an approximate LHY energy functional
(i.e., Petrov's prescription), as suggested by Petrov to artificially
remove an annoying, unphysical imaginary part in the energy functional
due to the lack of a consistent microscopic theory at that time \citep{Petrov2015}.
The density locking and the resulting single-mode description (i.e.,
an effective Gross-Pitaevskii equation) now have been the standard
theoretical framework to analyze and understand quantum droplets formed
in binary Bose mixtures \citep{Bottcher2021,Minardi2019}. However,
in a recent detailed analysis of the two experiments \citep{Cikojevic2020,Cikojevic2021},
Cikojevi\'{c} and collaborators found that a significant violation
of the density locking rule must be assumed in order to explain the
Barcelona experiment \citep{Cabrera2018}, where quantum droplets
were confined under a transverse trapping potential and a puzzling
low critical particle number for the formation of quantum droplets
was observed. For the Florence experiment, where quantum droplets
were created in free space without any confinement \citep{Semeghini2018},
instead there is no clear signal for the violation of the density
locking rule.

In this work, we aim to determining the optimal density ratio and
the allowed fluctuation in the density ratio, by taking advantage
of a recently developed microscopic bosonic pairing theory of quantum
droplets in binary Bose mixtures \citep{Hu2020a,Hu2020b}. This pairing
theory consistently removes the unphysical imaginary part due to the
softening complex Bogoliubov spectrum \citep{Hu2020a} and allows
us to explore a much wider parameter space away from the mean-field
collapse threshold at $a_{12}=a_{12}^{(c)}$, beyond the previous
works \citep{Staudinger2018,Ancilotto2018,Zin2021}. Remarkably, in
the case of unequal intra-species scattering lengths ($a_{11}\neq a_{22}$)
as in the experiments, we find a strong violation of the density locking
rule in the dense droplet regime, when the inter-species interaction
becomes very attractive. Our finding is consistent with the analysis
by Cikojevi\'{c} and coworkers \citep{Cikojevic2020,Cikojevic2021}
and therefore suggests a plausible way to explain the observed low
critical number for the droplet formation in the Barcelona experiment
\citep{Cabrera2018}.

The rest of the paper is organized in the following way. In the next
section (Sec. II), we briefly review the bosonic pairing theory of
attractive binary Bose mixtures. In Sec. III, we outline the procedure
of numerical calculations. In Sec. IV, we present the results on the
optimal density ratio and its fluctuation, for both equal and unequal
intra-species scattering lengths. The conclusions are drawn in Sec.
V. Finally, Appendix A is devoted to the stability analysis of our
bosonic pairing solutions and Appendix B discusses the allowed density
ratios based on the approximate LHY energy functional.

\section{The bosonic pairing theory of quantum droplets}

\subsection{The model Hamiltonian}

We consider a two-component Bose-Bose mixture in three dimensions,
such as the homonuclear $^{39}$K-$^{39}$K mixture investigated in
the two pioneering experiments \citep{Cabrera2018,Semeghini2018}.
Therefore, we take the same masses of the two bosonic components,
$m_{1}=m_{2}=m$. The bosonic atoms are interacting with each other
via some short-range interaction potentials, which can be well modeled
in the form of contact interaction potentials. By denoting the intra-species
interaction strengths as $g_{11}$ and $g_{22}$, and the inter-species
interaction strengths as $g_{12}=g_{21}$, the binary Bose mixture
in free space can then be described by the following model Hamiltonian
density, 
\begin{eqnarray}
\mathscr{H}\left(\mathbf{x}\right) & = & \mathscr{H}_{0}\left(\mathbf{x}\right)+\mathscr{H}_{\textrm{int}}\left(\mathbf{x}\right),\\
\mathscr{H}_{0}\left(\mathbf{x}\right) & = & \sum_{i=1,2}\phi_{i}^{\dagger}\left(\mathbf{x}\right)\left[-\frac{\hbar^{2}\nabla^{2}}{2m}-\mu_{i}\right]\phi_{i}\left(\mathbf{x}\right),\\
\mathscr{H}_{\textrm{int}}\left(\mathbf{x}\right) & = & \sum_{i,j=1,2}\frac{g_{ij}}{2}\phi_{i}^{\dagger}\left(\mathbf{x}\right)\phi_{j}^{\dagger}\left(\mathbf{x}\right)\phi_{j}\left(\mathbf{x}\right)\phi_{i}\left(\mathbf{x}\right).
\end{eqnarray}
Here, $\phi_{i}^{\dagger}(\mathbf{x})$ and $\phi_{i}(\mathbf{x})$
with $i=1,2$ are the creation and annihilation field operators of
the $i$th-component bosons and $\mu_{i}$ the corresponding chemical
potential.

\subsection{The pairing theory}

The bosonic pairing theory has been laid out in detail in previous
works \citep{Hu2020a,Hu2020b}. Here, we only briefly overview the
key ingredients of the theory. Inspired by the idea of Cooper pairing
between two fermions of unlike spins \citep{Hu2006,He2015,Hu2020e},
we assume that two bosons in different components may form a bosonic
pair due to the large inter-species attraction. Mathematically, such
a bosonic pairing can be conveniently described by introducing an
auxiliary pairing field $\Delta(x)$ in the context of functional
path-integral formalism, which, after the standard Hubbard--Stratonovich
transformation, turns the Hamiltonian density for inter-species interactions
into the following form: 
\begin{widetext}
\begin{equation}
\exp\left[-g_{12}\int dx\bar{\phi}_{1}\bar{\phi}_{2}\phi_{2}\phi_{1}\right]=\int\mathcal{D}\left[\Delta\left(x\right)\right]\exp\left\{ \int dx\left[\frac{\left|\Delta\left(x\right)\right|^{2}}{g_{12}}+\left(\bar{\Delta}\phi_{2}\phi_{1}+\bar{\phi}_{1}\bar{\phi}_{2}\Delta\right)\right]\right\} .
\end{equation}
Here, we have used the notations $x\equiv(\mathbf{x},\tau)$ and $\int dx\equiv\int d\mathbf{x}\int_{0}^{\beta}d\tau$,
and $\beta\equiv1/(k_{B}T)$ is the inverse temperature. The resulting
partition function, $\mathcal{Z}=\int\mathcal{D}[\phi_{1},\phi_{2}]e^{-\mathcal{S}},$
is then given by 
\begin{equation}
\mathcal{S}=\sum_{i=1,2}\int dx\left[\bar{\phi}_{i}\left(\partial_{\tau}-\frac{\hbar^{2}\nabla^{2}}{2m}-\mu_{i}\right)\phi_{i}+\frac{g_{ii}}{2}\bar{\phi}_{i}^{2}\phi_{i}^{2}\right]+\int dx\left[-\frac{\left|\Delta\left(x\right)\right|^{2}}{g_{12}}-\left(\bar{\Delta}\phi_{2}\phi_{1}+\bar{\phi}_{1}\bar{\phi}_{2}\Delta\right)\right],
\end{equation}
\end{widetext}

using which we can calculate the thermodynamic potential $\Omega=-k_{B}T\ln\mathcal{Z}$.
The pairing field does not have its own dynamics (i.e., as governed
by $\partial_{\tau}$ as the field operators $\phi_{i}$). As we consider
the thermodynamic limit, it often suffices to take a saddle-point
solution $\Delta(x)=\Delta$, which is both time and space independent.
At zero temperature, it is standard to take the Bogoliubov approximation
and assume that all bosons condense into the zero-momentum states
\citep{Pu1998}, and hence the bosonic field operators can be decomposed
as 
\begin{eqnarray}
\phi_{i}\left(x\right) & = & \phi_{ic}+\delta\phi_{i}\left(x\right),
\end{eqnarray}
with uniform condensate wave-functions $\phi_{ic}$ and fluctuations
$\delta\phi_{i}\left(x\right)$. The intra-species interaction terms
may then be approximated as 
\begin{equation}
\frac{g_{ii}}{2}\bar{\phi}_{i}^{2}\phi_{i}^{2}\simeq\frac{C_{i}^{2}}{2g_{ii}}+\frac{C_{i}}{2}\left(4\delta\bar{\phi}_{i}\delta\phi_{i}+\delta\bar{\phi}_{i}\delta\bar{\phi}_{i}+\delta\phi_{i}\delta\phi_{i}\right),
\end{equation}
where $C_{i}\equiv g_{ii}\phi_{ic}^{2}$. As a result, we find the
action 
\begin{equation}
\mathcal{S}\simeq\beta\Omega_{0}+\mathcal{S}_{B},
\end{equation}
where the condensate part $\Omega_{0}$ is given by 
\begin{equation}
\Omega_{0}=\sum_{i=1,2}\left(-\mu_{i}\phi_{ic}^{2}+\frac{C_{i}^{2}}{2g_{ii}}\right)-\frac{\Delta^{2}}{g_{12}}-2\Delta\phi_{1c}\phi_{2c},\label{eq:Omega0}
\end{equation}
and the fluctuation part by 
\begin{equation}
\mathcal{S}_{B}=\int dxdx'\bar{\Phi}\left(x\right)\left[-\mathscr{D}^{-1}\left(x,x'\right)\right]\Phi\left(x'\right)
\end{equation}
with the Nambu spinor defined as $\Phi(x)=[\delta\phi_{1}(x),\delta\bar{\phi}_{1}(x),\delta\phi_{2}(x),\delta\bar{\phi}_{2}(x)]^{T}$.
Here, the inverse Green function of bosons is given by 
\begin{widetext}
\begin{equation}
\mathscr{D}^{-1}\left(x,x'\right)=\left[\begin{array}{cccc}
-\partial_{\tau}-\hat{B}_{1} & -C_{1} & 0 & \Delta\\
-C_{1} & \partial_{\tau}-\hat{B}_{1} & \Delta & 0\\
0 & \Delta & -\partial_{\tau}-\hat{B}_{2} & -C_{2}\\
\Delta & 0 & -C_{2} & \partial_{\tau}-\hat{B}_{2}
\end{array}\right]\delta\left(x-x'\right),
\end{equation}
where $\hat{B}_{i}(x)\equiv-\hbar^{2}\nabla^{2}/(2m)-\mu_{i}+2C_{i}$.
Due to the presence of the delta function $\delta\left(x-x'\right)$,
it is convenient to work in momentum space by taking a Fourier transform.
After replacing $-\partial_{\tau}$ with the bosonic Matasubara frequencies
$i\omega_{m}$ (i.e., $\omega_{m}=2\pi mk_{B}T$ with $m\subseteq\mathbb{Z}$),
performing the analytic continuation $i\omega_{m}\rightarrow\omega+i0^{+}$
(i.e., $-\partial_{\tau}\rightarrow\omega+i0^{+}$) and taking the
replacement $\hat{B}_{i}\rightarrow B_{i\mathbf{k}}=\varepsilon_{\mathbf{k}}-\mu_{i}+2C_{i}$
with $\varepsilon_{\mathbf{k}}=\hbar^{2}\mathbf{k}^{2}/(2m)$, it
is straightforward to write down $\mathscr{D}^{-1}(\mathbf{k},\omega)$.
By solving the poles of the bosonic Green function, i.e., $\det[\mathscr{D}^{-1}(\mathbf{k},\omega\rightarrow E(\mathbf{k}))]=0$,
or more explicitly, 
\begin{equation}
\omega^{4}-\omega^{2}\left[\left(B_{1\mathbf{k}}^{2}-C_{1}^{2}\right)+\left(B_{2\mathbf{k}}^{2}-C_{2}^{2}\right)-2\Delta^{2}\right]+\left[\left(B_{1\mathbf{k}}^{2}-C_{1}^{2}\right)\left(B_{2\mathbf{k}}^{2}-C_{2}^{2}\right)-2\left(B_{1\mathbf{k}}B_{2\mathbf{k}}+C_{1}C_{2}\right)\Delta^{2}+\Delta^{4}\right]=0,\label{eq:Det}
\end{equation}
we obtain the two Bogoliubov dispersion relations, 
\begin{equation}
E_{\pm}^{2}\left(\mathbf{k}\right)=\left[\mathcal{A}_{+}\left(\mathbf{k}\right)-\Delta^{2}\right]\pm\sqrt{\mathcal{A}_{-}^{2}\left(\mathbf{k}\right)+\Delta^{2}\left[\left(C_{1}+C_{2}\right)^{2}-\left(B_{1\mathbf{k}}-B_{2\mathbf{k}}\right)^{2}\right]},
\end{equation}
\end{widetext}

where we have defined 
\begin{equation}
\mathcal{A}_{\pm}\left(\mathbf{k}\right)=\frac{\left(B_{1\mathbf{k}}^{2}-C_{1}^{2}\right)\pm\left(B_{2\mathbf{k}}^{2}-C_{2}^{2}\right)}{2}.
\end{equation}

\subsection{Thermodynamic potential}

By taking the derivative of the condensate thermodynamic potential
$\Omega_{0}$ in Eq. (\ref{eq:Omega0}) with respect to $\phi_{1c}$
and $\phi_{2c}$ \citep{Salasnich2016}, we obtain 
\begin{eqnarray}
-\mu_{1}\phi_{1c}+g_{11}\phi_{1c}^{3}-\Delta\phi_{2c} & = & 0,\\
-\mu_{2}\phi_{2c}+g_{22}\phi_{2c}^{3}-\Delta\phi_{1c} & = & 0,
\end{eqnarray}
from which, we have 
\begin{eqnarray}
C_{1} & = & \mu_{1}+\Delta\frac{\phi_{2c}}{\phi_{1c}},\\
C_{2} & = & \mu_{2}+\Delta\frac{\phi_{1c}}{\phi_{2c}}.
\end{eqnarray}
On the other hand, it is straightforward to rewrite the condensate
thermodynamic potential in the form, 
\begin{equation}
\Omega_{0}=-\frac{\Delta^{2}}{g_{12}}-\frac{C_{1}^{2}}{2g_{11}}-\frac{C_{2}^{2}}{2g_{22}}.
\end{equation}
The action for quantum fluctuations $\mathcal{S}_{B}$ has a bilinear
form. It provides the following well-known LHY contribution to the
thermodynamic potential at zero temperature \citep{Hu2020c,Salasnich2016},
\begin{align}
\Omega_{\textrm{LHY}} & =\frac{k_{B}T}{2}\sum_{\mathbf{q},i\omega_{m}}\ln\det\left[\mathscr{-D}^{-1}\left(\mathbf{q},i\omega_{m}\right)\right]e^{i\omega_{m}0^{+}},\\
 & =\frac{1}{2}\sum_{\mathbf{k}}\left[E_{+}\left(\mathbf{k}\right)+E_{-}\left(\mathbf{k}\right)-B_{1\mathbf{k}}-B_{2\mathbf{k}}\right].
\end{align}
It is worth noting that both $\Omega_{0}$ and $\Omega_{\textrm{LHY}}$
have ultraviolet divergence, as a well-known consequence of using
the contact inter-particle interaction potentials. This however can
be cured by regularizing the bare interaction strengths $g_{ij}$
in terms of the $s$-wave scattering lengths $a_{ij}$ \citep{Salasnich2016}:
$g_{ij}^{-1}=m/(4\pi\hbar^{2}a_{ij})-\mathcal{V}^{-1}\sum_{\mathbf{k}}m/(\hbar^{2}\mathbf{k}^{2})$.
To make the notations simpler, hereafter we ignore the volume $\mathcal{V}$
in equations. In other words, informally we always set the volume
$\mathcal{V}=1$. After this regularization, the divergences in $\Omega_{0}$
and $\Omega_{\textrm{LHY}}$ exactly cancel with each other. We finally
obtain the total thermodynamic potential within the Bogoliubov approximation
as 
\begin{widetext}
\begin{equation}
\Omega\left(\mu_{1},\mu_{2},\Delta\right)=-\frac{m}{4\pi\hbar^{2}}\left[\frac{\Delta^{2}}{a_{12}}+\frac{C_{1}^{2}}{2a_{11}}+\frac{C_{2}^{2}}{2a_{22}}\right]+\frac{1}{2}\sum_{\mathbf{k}}\left[E_{+}\left(\mathbf{k}\right)+E_{-}\left(\mathbf{k}\right)-B_{1\mathbf{k}}-B_{2\mathbf{k}}+\frac{C_{1}^{2}+C_{2}^{2}+2\Delta^{2}}{\hbar^{2}\mathbf{k}^{2}/m}\right].
\end{equation}
\end{widetext}

From now on, we refer to the \emph{regularized} LHY term, i.e., the
second term on the right-hand side of the above equation, as the LHY
thermodynamic potential \citep{Salasnich2016}. As in the standard
Bardeen--Cooper--Schrieffer (BCS) theory of fermionic superfluidity,
for the given chemical potentials $\mu_{1}$ and $\mu_{2}$, we need
to minimize $\Omega(\mu_{1},\mu_{2},\Delta)$ with respect to the
pairing parameter at $\Delta=\Delta_{0}$. By denoting $\Omega(\mu_{1},\mu_{2})=\Omega(\mu_{1},\mu_{2},\Delta_{0})$,
in turn, we need to tune the two chemical potentials to satisfy the
two number equations, i.e., 
\begin{equation}
n_{i}=-\frac{\partial\Omega\left(\mu_{1},\mu_{2}\right)}{\partial\mu_{i}}\,,
\end{equation}
for both components $i=1,2$. According to the Bogoliubov approximation,
we also have the two conditions $\phi_{ic}=\sqrt{n_{i}}$.

\section{Numerical calculations}

The procedure of our numerical calculations is straightforward. A
crucial step is to minimize the thermodynamic potential $\Omega$
with respect to the pairing parameter $\Delta$. For this purpose,
let us introduce the dimensionless variables, $t\equiv\hbar^{2}\mathbf{k}^{2}/(4m\Delta)$,
$\tilde{\mu}_{i}\equiv\mu_{i}/(2\Delta)$, $\tilde{C}_{i}\equiv C_{i}/(2\Delta)$,
and $\tilde{B}_{it}=B_{i\mathbf{k}}/(2\Delta)$. Let us also denote
$r\equiv\phi_{2c}/\phi_{1c}=\sqrt{n_{2}/n_{1}}$. Explicitly, we find
that, 
\begin{eqnarray}
\tilde{C}_{1} & = & \tilde{\mu}_{1}+r/2,\\
\tilde{C}_{2} & = & \tilde{\mu}_{2}+\left(2r\right)^{-1},\\
\tilde{B}_{1t} & = & t+\tilde{\mu}_{1}+r,\\
\tilde{B}_{2t} & = & t+\tilde{\mu}_{2}+r^{-1}.
\end{eqnarray}
Using these notations, we can cast $\Omega_{\textrm{LHY}}$ into the
form, 
\begin{equation}
\Omega_{\textrm{LHY}}=\frac{32\sqrt{2}m^{3/2}}{15\pi^{2}\hbar^{3}}\Delta^{5/2}f\left(\tilde{\mu}_{1},\tilde{\mu}_{2}\right),
\end{equation}
where the function $f(\tilde{\mu}_{1},\tilde{\mu}_{2})$ is defined
by 
\begin{widetext}
\begin{equation}
f\left(\tilde{\mu}_{1},\tilde{\mu}_{2}\right)=\frac{15\sqrt{2}}{32}\int_{0}^{\infty}dt\sqrt{t}\left[\tilde{E}_{+}\left(t\right)+\tilde{E}_{-}\left(t\right)-\left(\tilde{B}_{1t}+\tilde{B}_{2t}\right)+\frac{\tilde{C}_{1}^{2}+\tilde{C}_{2}^{2}+1/2}{2t}\right],
\end{equation}
with 
\begin{equation}
\tilde{E}_{\pm}^{2}\left(t\right)=\frac{\left(\tilde{B}_{1t}^{2}-\tilde{C}_{1}^{2}\right)+\left(\tilde{B}_{2t}^{2}-\tilde{C}_{2}^{2}\right)}{2}-\frac{1}{4}\pm\frac{1}{2}\sqrt{\left[\left(\tilde{B}_{1t}^{2}-\tilde{C}_{1}^{2}\right)-\left(\tilde{B}_{2t}^{2}-\tilde{C}_{2}^{2}\right)\right]^{2}+\left(\tilde{C}_{1}+\tilde{C}_{2}\right)^{2}-\left(\tilde{B}_{1t}-\tilde{B}_{2t}\right)^{2}}.
\end{equation}
The total thermodynamic potential is then given by 
\begin{equation}
\Omega\left(\mu_{1},\mu_{2},\Delta\right)\simeq-\frac{m\Delta^{2}}{2\pi\hbar^{2}}\left[\frac{\tilde{C}_{1}^{2}}{a_{11}}+\frac{\tilde{C}_{2}^{2}}{a_{22}}+\frac{1}{2a_{12}}\right]+\frac{32\sqrt{2}m^{3/2}}{15\pi^{2}\hbar^{3}}\Delta^{5/2}f\left(\tilde{\mu}_{1},\tilde{\mu}_{2}\right).\label{eq:omega}
\end{equation}
Numerically, it is convenient to take $\hbar^{2}/(2ma_{11}^{2})$
and $\hbar^{2}/(2ma_{11}^{5})$ as the units for energy (i.e., for
the chemical potentials, pairing parameter and Bogoliubov spectra)
and for thermodynamic potential, respectively. By denoting $\bar{\Delta}\equiv2ma_{11}^{2}\Delta/\hbar^{2}$
and $\bar{\Omega}\equiv2ma_{11}^{5}\Omega/\hbar^{2}$, we end up with
a dimensionless thermodynamic potential, 
\begin{equation}
\bar{\Omega}\left(\tilde{\mu}_{1},\tilde{\mu}_{2},\bar{\Delta}\right)=-\frac{\bar{\Delta}^{2}}{4\pi^{2}}\left[\tilde{C}_{1}^{2}+\frac{a_{11}}{a_{22}}\tilde{C}_{2}^{2}+\frac{a_{11}}{2a_{12}}\right]+\frac{16\bar{\Delta}^{5/2}}{15\pi^{2}}f\left(\tilde{\mu}_{1},\tilde{\mu}_{2}\right).
\end{equation}
\end{widetext}

For small chemical potentials $\tilde{\mu}_{1},\tilde{\mu}_{2}\sim0$,
which are typically achieved for a quantum droplet state, both the
square bracket and the function $f(\tilde{\mu}_{1},\tilde{\mu}_{2})$
in the above equation are positive. The balance of these two terms,
i.e., the attractive mean-field contribution and the repulsive LHY
thermodynamic potential, ensures the existence of a global minimum
in the total thermodynamic potential as a function of the pairing
parameter.

\begin{figure}[t]
\begin{centering}
\includegraphics[width=0.5\textwidth]{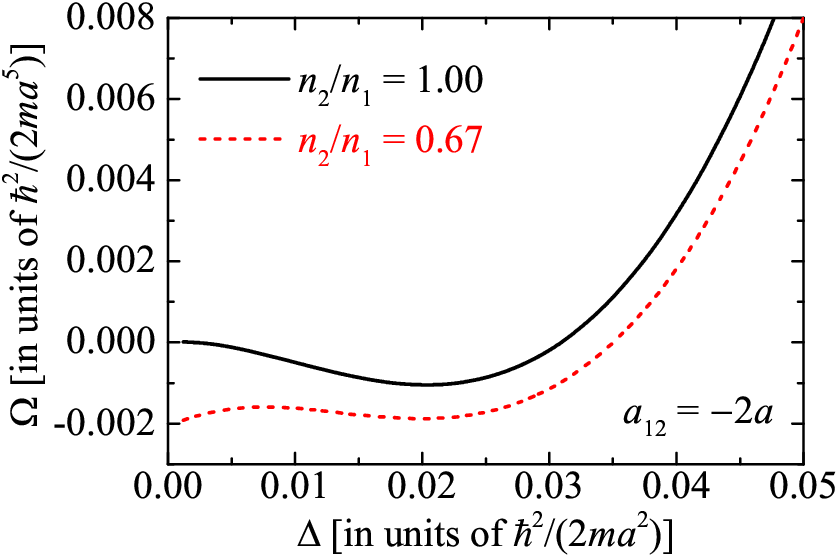} 
\par\end{centering}
\caption{\label{fig1_omega} Thermodynamic potential $\Omega$, in units of
$\hbar^{2}/(2ma^{5})$, as a function of the pairing parameter $\Delta$,
at two different density ratios $n_{2}/n_{1}=1$ (black solid line)
and $n_{2}/n_{1}=0.67$ (red dashed line). We take equal intra-species
scattering lengths $a_{11}=a_{22}=a$ and an inter-species scattering
length $a_{12}=-2a$. At this strongly attractive inter-species interaction,
we consider a large total density, as given by a large gas parameter
$na^{3}=0.0012$. The pairing parameter $\Delta$ is measured in units
of $\hbar^{2}/(2ma^{2})$. The two chemical potentials of the different
components are determined by satisfying the relevant number equations.}
\end{figure}

This is illustrated in Fig. \ref{fig1_omega} by the solid black line,
for the case of equal populations of bosons in the two components
($n_{1}=n_{2}$) and equal intra-species interaction strengths ($a_{11}=a_{22}=a$),
where we clearly identify a global minimum at $\Delta\simeq0.02\hbar^{2}/(2ma$$^{2}$)
at the inter-species scattering length $a_{12}=-2a$. A population
imbalance may make the chemical potentials $\mu_{i}$ to significantly
deviate from zero. As a consequence, with increasing population imbalance
(i.e., $n_{2}/n_{1}\rightarrow0$ or $\infty$) the global minimum
in the total thermodynamic potential gradually turns into a local
minimum (an example is given by the red dashed line for $n_{2}=0.67n_{1}$
in Fig. \ref{fig1_omega}) and eventually disappears.

Therefore, for the existence of quantum droplets, there is a parameter
window for the density ratio $n_{2}/n_{1}$. Furthermore, with increasing
total density $n=n_{1}+n_{2}$, a quantum droplet state is unstable
towards a gas phase \citep{Petrov2015}, so the parameter window of
the allowed density ratio $n_{2}/n_{1}$ would become narrower. As
we shall see, the window will eventually shrink to a single point,
which we refer to as the optimal density ratio. A quantum droplet
at this optimal density ratio is most robust against the change in
total density.

\section{Results and discussions}

Our goal is to find the parameter regime for the existence of a stable
self-bound droplet in free space, i.e., a droplet in equilibrium with
vacuum. Thus, we should use the following criteria: (1) the pressure
$P=0$, and (2) the chemical potentials $\mu_{i}\le0$.

\subsection{Equal intra-species interactions}

To better understand the above picture, let us start with equal intra-species
interactions. In this case, the optimal density ratio must be $n_{2}/n_{1}=1$,
so the two bosonic components are symmetric and the resulting quantum
droplet state would be the most robust. However, the density ratio
could fluctuate within a certain interval, due to the non-negligible
chemical potentials.

\begin{figure}[t]
\begin{centering}
\includegraphics[width=0.5\textwidth]{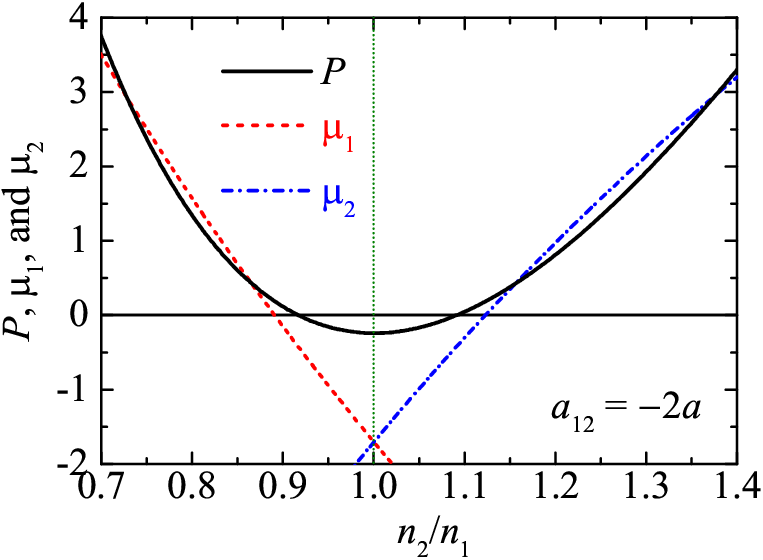} 
\par\end{centering}
\caption{\label{fig2_pressure} Pressure $P$ (in units of $10^{-7}\hbar^{2}/(2ma^{5})$)
and the two chemical potentials $\mu_{1}$ and $\mu_{2}$ (in units
of $10^{-3}\hbar^{2}/(2ma^{2})$), as a function of the density ratio
$n_{2}/n_{1}$. We consider equal intra-species scattering lengths
$a_{11}=a_{22}=a$ and an inter-species scattering length $a_{12}=-2a$.
The total density is given by the gas parameter $na^{3}=8.8\times10^{-4}$.
The vertical green dotted line indicates the optimal density ratio
$n_{2}/n_{1}=1$.}
\end{figure}

To see this, in Fig. \ref{fig2_pressure} we show the pressure $P=-\Omega$
of the binary mixture as a function of the density ratio $n_{2}/n_{1}$
by a black solid line, for a specific inter-species scattering length
$a_{12}=-2a$. We choose a total density that corresponds to a gas
parameter $na^{3}=8.8\times10^{-4}$. At this density, the pressure
is negative at the balanced component density $n_{1}=n_{2}$. With
increasing or decreasing density ratio away from the population balance,
the pressure always increases and becomes zero at the critical density
ratios $(n_{2}/n_{1})_{P=0}^{(L)}<1$ and $(n_{2}/n_{1})_{P=0}^{(R)}>1$,
respectively. As the two components are symmetric due to the equal
intra-species scattering lengths, we have a useful inverse relation,
\begin{equation}
\left(\frac{n_{2}}{n_{1}}\right)_{P=0}^{(L)}=\left[\left(\frac{n_{2}}{n_{1}}\right)_{P=0}^{(R)}\right]^{-1}.
\end{equation}
The binary mixture at these density ratios is a self-bound quantum
droplet, if the chemical potentials of the two components are smaller
than zero (i.e., the chemical potential of the surrounding vacuum)
\citep{Petrov2015,Hu2020a}. In Fig. \ref{fig2_pressure}, we also
show the chemical potentials by using the red dashed line (for $\mu_{1}$)
and the blue dot-dashed line (for $\mu_{2}$), which become negative
once the density ratio is larger or smaller than the threshold $(n_{2}/n_{1})_{\mu_{1}=0}<1$
or $(n_{2}/n_{1})_{\mu_{2}=0}=[(n_{2}/n_{1})_{\mu_{1}=0}]^{-1}>1$,
respectively. We find that the mixture indeed has negative chemical
potentials at both critical density ratios for zero pressure. Hence,
we could obtain a quantum droplet state with a density ratio different
from the optimal value, which is $n_{2}/n_{1}=1$ in this case.

\begin{figure}[t]
\begin{centering}
\includegraphics[width=0.5\textwidth]{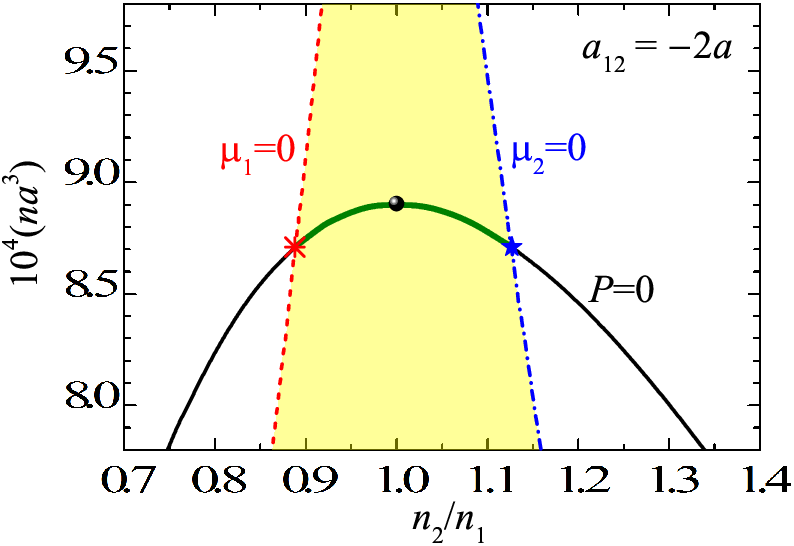} 
\par\end{centering}
\caption{\label{fig3_n2n1_a12fix} The zero pressure line (the solid line for
$(n_{2}/n_{1})_{P=0}^{(L)}$ or $(n_{2}/n_{1})_{P=0}^{(R)}$) and
the zero chemical potential lines (the red dashed line for $(n_{2}/n_{1})_{\mu_{1}=0}$
and the blue dot-dashed line for $(n_{2}/n_{1})_{\mu_{2}=0}$), in
the plane of the density ratio ($n_{2}/n_{1}$) and the total density
($na^{3}$). We consider equal intra-species scattering lengths $a_{11}=a_{22}=a$
and set an inter-species scattering length $a_{12}=-2a$. The yellow
area marks the stable parameter space, in which the binary mixture
will eventually turn into a quantum droplet, with a density ratio
located at the green line between the two asterisks (i.e, the red
eight spoked asterisk and the blue star). The black circle indicates
the optimal density ratio $n_{2}/n_{1}=1$, for equal intra-species
scattering lengths.}
\end{figure}

In Fig. \ref{fig3_n2n1_a12fix}, we report the critical density ratios
$(n_{2}/n_{1})_{P=0}^{(L)}$ or $(n_{2}/n_{1})_{P=0}^{(R)}$ for zero
pressure and the threshold density ratios $(n_{2}/n_{1})_{\mu_{1}=0}$
and $(n_{2}/n_{1})_{\mu_{2}=0}$ for zero chemical potentials, at
varying total densities, which form the zero pressure line (i.e.,
the black solid line) and the two zero chemical potential lines (the
red dashed line where $\mu_{1}=0$ and the blue dot-dashed line where
$\mu_{2}=0$). As the total density increases, the two critical density
ratios for zero pressure smoothly connect at the optimal density ratio
$n_{2}/n_{1}=1$ (see the black circle), as expected. Importantly,
this zero pressure line crosses the two zero chemical potential lines
at the two points, which are marked in the figure by the red eight
spoked asterisk and the blue star, respectively.

We now would like to claim that the zero pressure line between these
two asterisks, colored in green in the figure, denotes the parameter
window of the \emph{fluctuated} density ratio, at which a self-bound
quantum droplet with zero pressure is allowed to maintain. Moreover,
the area between the two zero chemical potential lines, which is highlighted
in the figure in yellow, indicates the \emph{stable} region for the
binary Bose mixture to eventually become a quantum droplet \citep{Staudinger2018,Zin2021}.
To understand it, let us assume that initially the mixture is in the
yellow area above the green line, where $P>0$. As its pressure is
positive, the mixture will expand to reduce its density and thereby
move downwards. It will eventually touch the green line to become
a self-bound quantum droplet. Instead, if initially the mixture is
in the yellow area below the green line, it will shrink and increase
its density because of the negative pressure, and eventually reach
the green line. In the latter situation, of course, the mixture may
also first touch the zero chemical potential lines. In this case,
it will self-evaporate its excess component to increase (if the red
line for $\mu_{1}=0$ is touched) or reduce (if the blue dot-dashed
line for $\mu_{2}=0$ is touched) the density ratio \citep{Petrov2015}.
The mixture will then move towards inside and eventually fall onto
the green line to create a stable quantum droplet.

In the above analysis, we do not explicitly discuss the mechanical
instability (for collapse) and the spinodal instability (for phase
separation of the two components) of the binary mixture, as recently
emphasized in Refs. \citep{Staudinger2018,Ancilotto2018,Zin2021}.
This is because, in our bosonic pairing theory the curvature Hessian
matrix is always positively defined, once a pairing solution is available.
Therefore, we do not find any instabilities in the yellow area of
Fig. \ref{fig3_n2n1_a12fix} that we are interested in. For more details,
please see Appendix A.

\begin{figure}[t]
\begin{centering}
\includegraphics[width=0.5\textwidth]{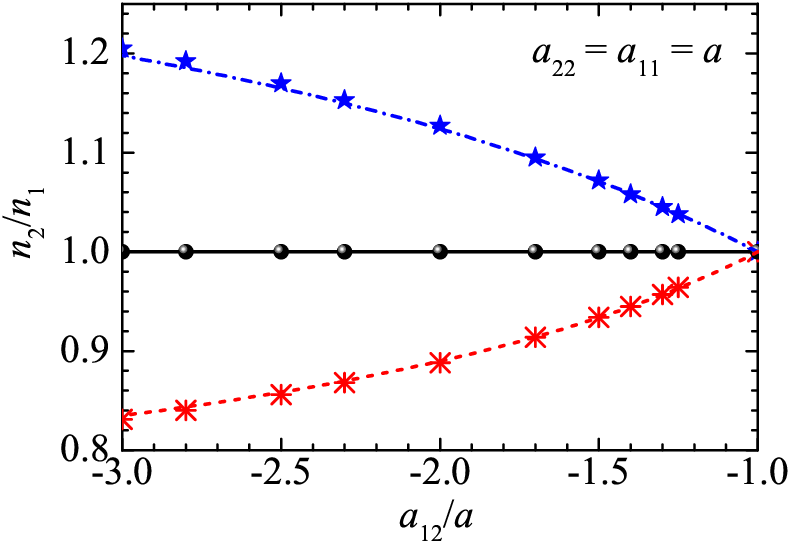} 
\par\end{centering}
\caption{\label{fig4_n2n1_a12dep} The optimal density ratio (black circles)
and the allowed fluctuation of the density ratio (i.e., the upper
bound by blue stars and the lower bound by red eight spoked asterisks),
as a function of the inter-species scattering length, in the case
of equal intra-species scattering lengths $a_{11}=a_{22}=a$. The
red dashed line and blue dot-dashed line are the theoretical predictions
from Ref. \citep{Zin2021} (see the discussions in Sec. IVC).}
\end{figure}

In Fig. \ref{fig4_n2n1_a12dep}, we investigate the dependence of
the green line on the inter-species scattering length. A stable self-bound
quantum droplet could be created, for any density ratio lies between
the two lines connecting either the red eight spoked asterisks or
the blue stars. Near the threshold of mean-field collapse, we find
that the density ratio can only change just by a few percent (i.e.,
about $3\%$ at $a_{12}=-1.2a$) away from the optimal density ratio,
consistent with the previous analyses \citep{Staudinger2018,Ancilotto2018,Zin2021}.
However, as the inter-species scattering length becomes more attractive,
the allowed fluctuation in the density ratio becomes larger. For example,
our microscopic pairing theory predicts a more than $10\%$ variation
in the density ratio at $a_{12}=-2a$, which should be sizable for
experimental examination.

\begin{figure}[t]
\begin{centering}
\includegraphics[width=0.5\textwidth]{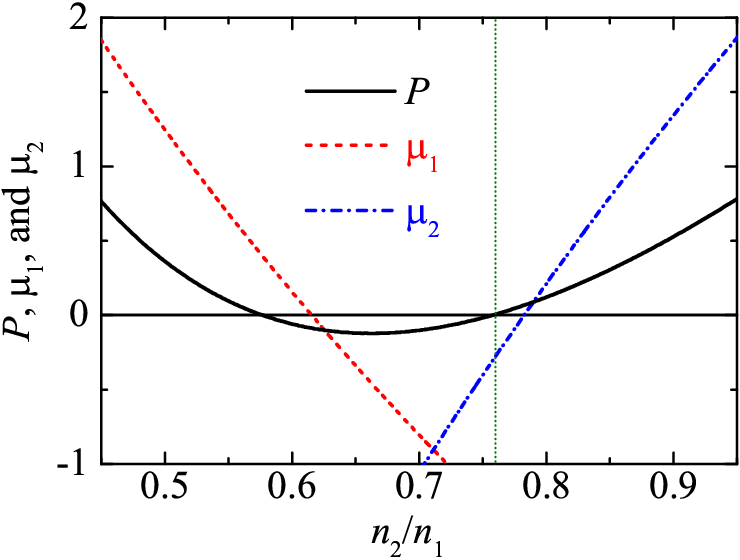} 
\par\end{centering}
\caption{\label{fig5_pressure} Pressure $P$ (in units of $10^{-7}\hbar^{2}/(2ma_{11}^{5})$)
and the two chemical potentials $\mu_{1}$ and $\mu_{2}$ (in units
of $10^{-3}\hbar^{2}/(2ma_{11}^{2})$), as a function of the density
ratio $n_{2}/n_{1}$. We consider unequal intra-species scattering
lengths with a ratio $a_{11}/a_{22}\simeq0.577$ and an inter-species
scattering length $a_{12}=-2a$, where $a=\sqrt{a_{11}a_{22}}$. The
total density corresponds to a gas parameter $na_{11}^{3}=3.5\times10^{-4}$.
The vertical green dotted line indicates the optimal density ratio
predicted by Eq. (\ref{eq:DensityRatioPetrov}), i.e., $n_{2}/n_{1}=\sqrt{a_{11}/a_{22}}\simeq0.76$.}
\end{figure}

\subsection{Unequal intra-species interactions}

We now turn to consider the more general and interesting case of unequal
intra-species scattering lengths. In this case, it is illustrative
to briefly review how to obtain the density locking rule Eq. (\ref{eq:DensityRatioPetrov})
from the bosonic pairing theory \citep{Hu2020a}. Near the threshold
of mean-field collapse, where the chemical potentials $\mu_{i}\sim0$,
we may expand the thermodynamic potential in Eq.~(\ref{eq:omega})
in powers of $\mu_{1}$ and $\mu_{2}$ as 
\begin{equation}
\Omega\left(\mu_{1},\mu_{2},\Delta\right)=\Omega\left(0,0,\Delta\right)-\frac{m\Delta}{4\pi\hbar^{2}}\left[\frac{r\mu_{1}}{a_{11}}+\frac{\mu_{2}}{a_{22}r}\right],\label{eq:omegaAPPROX}
\end{equation}
where we have neglected terms of the order $\mathcal{O}\left(\mu_{1}^{2},\mu_{2}^{2}\right)$
or higher. Taking the derivative of $\Omega(\mu_{1},\mu_{2},\Delta)$
with respect to $\mu_{1}$ and $\mu_{2}$ and setting the saddle point
$\Delta=\Delta_{0}$, we find that 
\begin{eqnarray}
n_{1} & = & \frac{m\Delta_{0}r}{4\pi\hbar^{2}a_{11}},\\
n_{2} & = & \frac{m\Delta_{0}}{4\pi\hbar^{2}a_{22}r}.
\end{eqnarray}
By dividing these two expressions with each other, we obtain the optimal
density ratio 
\begin{equation}
r^{2}=\frac{n_{2}}{n_{1}}=\sqrt{\frac{a_{11}}{a_{22}}}\,,\label{38}
\end{equation}
which coincides with Petrov's prediction in Eq.~(\ref{eq:DensityRatioPetrov}).
However, we stress that Eq.~(\ref{38}) is only valid near the threshold
of mean-field collapse. As we will show below, away from the threshold,
both chemical potentials $\mu_{1}$ and $\mu_{2}$ may become significant
(i.e., compared with the pairing parameter $\Delta$). The approximate
form of the thermodynamic potential in Eq. (\ref{eq:omegaAPPROX})
is no longer valid. As a consequence, we may find an optimal density
ratio significantly different from Eq.~(\ref{38}).

\begin{figure}[t]
\begin{centering}
\includegraphics[width=0.5\textwidth]{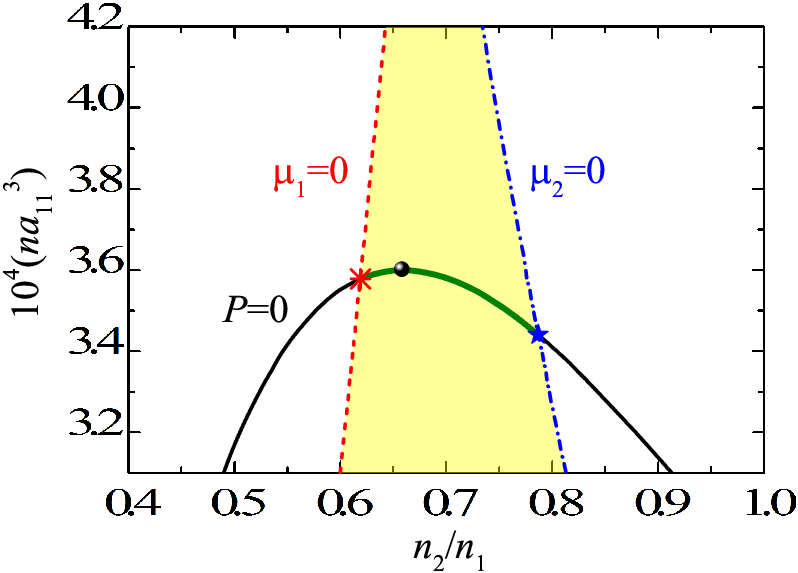} 
\par\end{centering}
\caption{\label{fig6_n2n1_a12fix} The zero pressure line (the solid line)
and the zero chemical potential lines (the red dashed line and the
blue dot-dashed line), in the plane of the density ratio ($n_{2}/n_{1}$)
and the total density ($na_{11}^{3}$), in the case of unequal intra-species
scattering lengths $a_{11}/a_{22}\simeq0.577$. We set an inter-species
scattering length $a_{12}=-2a$, where $a=\sqrt{a_{11}a_{22}}$. The
yellow area marks the stable parameter space, in which the binary
mixture will eventually turn into a quantum droplet located at the
green line between the two asterisks. The black circle indicates the
optimal density ratio, which is very different from the predicted
value of $n_{2}/n_{1}\simeq0.76$.}
\end{figure}

As a concrete example, in Fig. \ref{fig5_pressure} we consider an
unequal intra-species scattering lengths with $a_{11}/a_{22}\simeq0.577$
and plot the pressure as a function of the density ratio, along with
the two chemical potentials. We choose the total density $na_{11}^{3}=3.5\times10^{-4}$,
so the minimum pressure is negative. An immediate observation is that
the minimum pressure is located at a density ratio, which is clearly
different from the one predicted by Eq.~(\ref{38}), i.e., $n_{2}/n_{1}=\sqrt{a_{11}/a_{22}}\simeq0.76$
(indicated by the vertical green dotted line in Fig. \ref{fig5_pressure}).
The two chemical potentials also cross at a certain density ratio,
below the predicted optimal density ratio. Overall, the inverse symmetry
between the two critical density ratios for zero pressure and between
the two threshold density ratios for zero chemical potentials, as
mirrored by the optimal density ratio in Fig. \ref{fig2_pressure},
are completely lost, because of the unequal intra-species interactions.

In Fig. \ref{fig6_n2n1_a12fix}, we show the critical density ratios
and the threshold density ratios, as a function of the total density
$n$, keeping the ratio of the two intra-species scattering lengths
$a_{11}/a_{22}\simeq0.577$. As in the case of equal intra-species
interactions, the two critical density ratios $(n_{2}/n_{1})_{P=0}^{(L)}$
and $(n_{2}/n_{1})_{P=0}^{(R)}$ for zero pressure smoothly connect
at a point (see the black circle), which defines the optimal density
ratio at the highest total density for creating a self-bound quantum
droplet. The two crossing points of the two zero chemical potential
lines and the zero pressure line, indicated respectively by the two
asterisks, similarly mark a parameter window of the density ratio
for a stable quantum droplet. We observe that the two asterisks distribute
highly asymmetrically around the optimal density ratio (i.e., the
black circle). We note that, near the threshold of the mean-field
collapse, a similar asymmetric distribution for the allowed density
ratio was previously found in a stability analysis \citep{Zin2021},
based on the approximate LHY energy functional.

\begin{figure}[t]
\begin{centering}
\includegraphics[width=0.5\textwidth]{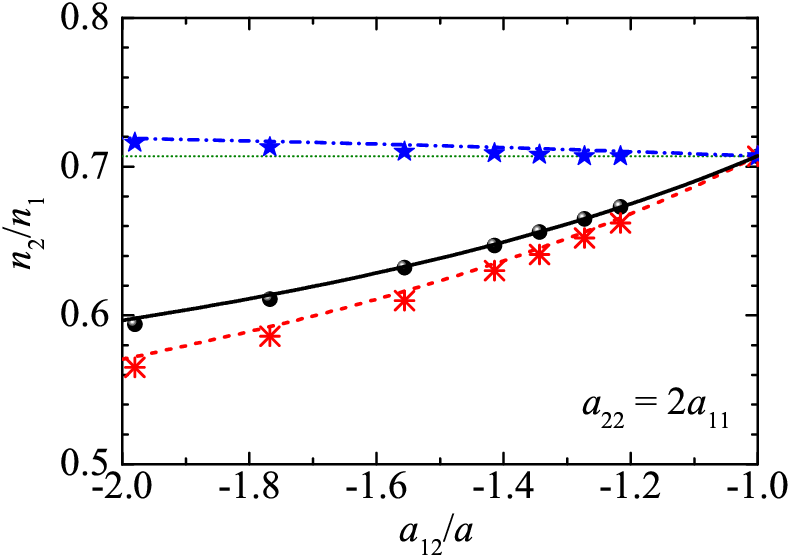} 
\par\end{centering}
\caption{\label{fig7_n2n1_a12dep} The optimal density ratio (black circles)
and the allowed fluctuation of the density ratio (i.e., the upper
bound by blue stars and the lower bound by red eight spoked asterisks),
as a function of the inter-species scattering length, in the case
of unequal intra-species scattering lengths $a_{11}/a_{22}=1/2$.
The horizontal green dotted line indicates the optimal density ratio
predicted by Eq. (\ref{eq:DensityRatioPetrov}), i.e., $n_{2}/n_{1}=\sqrt{a_{11}/a_{22}}\simeq0.707$.
The others lines are the theoretical predictions from Ref. \citep{Zin2021}
by using the approximate LHY theory (see the discussions in Sec. IVC).
Despite the similar results in the density ratios obtained by the
approximate LHY theory and by the bosonic pairing theory, the corresponding
total densities are dramatically different (i.e., by a factor of about
$5$). We refer to Appendix B for more details.}
\end{figure}

Finally, in Fig. \ref{fig7_n2n1_a12dep} we report the optimal density
ratio as a function of the inter-species scattering length, together
with the upper and lower bounds of the allowed fluctuation in density
ratio (i.e., the two asterisks). Here, we slightly change the ratio
of the two intra-species scattering lengths to $a_{11}/a_{22}=0.5$,
so the predicted optimal density ratio in Eq. (\ref{eq:DensityRatioPetrov})
is given by $n_{2}/n_{1}\simeq0.707$. It is readily seen that, with
increasing inter-species attraction, our calculated optimal density
ratio (i.e., black circles) deviates more significantly from the predicted
optimal density ratio. Interestingly, at this ratio of intra-species
scattering lengths, the upper bound of the allowed density ratio (i.e.,
the blue stars) roughly follows the predicted optimal density ratio,
while the lower bound of the allowed density ratio (i.e., the red
eight spoked asterisks) does not differ too much with our calculated
optimal density ratio.

\subsection{Comparison with the previous works}

In the previous works \citep{Staudinger2018,Ancilotto2018,Zin2021},
the stability of the droplet state with respect to a varying density
ratio was analyzed. In particular, based on the approximate LHY energy
functional, Zin \textit{et al}. derived some analytic expressions
to determine the zero pressure line and the zero chemical potential
lines, see, e.g., Eqs. (24)-(26) in Ref. \citep{Zin2021}. By using
these equations, the parameter window of allowed density ratios close
to the threshold of mean-field collapse was discussed \citep{Zin2021}.

We have applied their equations to calculate the threshold density
ratios and to determine the optimal density ratio far away from the
mean-field collapse threshold. For more details, we refer to Appendix
B. The results are shown in Fig. \ref{fig4_n2n1_a12dep} and Fig.
\ref{fig7_n2n1_a12dep} for the cases of equal and unequal intra-species
interactions, respectively. We find an excellent agreement between
the predictions from the approximate LHY energy functional theory
and our bosonic pairing theory, for all the inter-species scattering
lengths under investigation. This is a remarkable coincidence, considering
the fact that the the approximate LHY theory is only applicable near
the threshold of mean-field collapse, a restriction that is strictly
followed by all the previous stability studies \citep{Ancilotto2018,Zin2021}.
This coincidence indicates the energy functional of the bosonic pairing
theory may have a similar structure as the approximate LHY theory,
upon changing the relative densities of the two components. Therefore,
both theories predict the similar results for the allowed density
ratios.

However, we would like to emphasize that such coincidence does not
imply that the bosonic pairing theory is equivalent to the approximate
LHY theory. Actually the two theories predict very different total
equilibrium density away from the threshold of mean-field collapse.
For example, at $a_{12}=-2a$ and at the optimal density ratio shown
in Fig. \ref{fig7_n2n1_a12dep}, the equilibrium total density predicted
by our bosonic pairing theory is more than five times smaller than
the LHY equilibrium density (see Appendix B). This reduction in the
equilibrium density was discussed in detail in Ref. \citep{Hu2020a},
benchmarked with the \textit{ab inito} quantum Monte Carlo data \citep{Cikojevic2019}.

\subsection{Experimental relevance}

For the homonuclear $^{39}$K-$^{39}$K mixture used in the two experiments
\citep{Cabrera2018,Semeghini2018}, the intra-species scattering length
$a_{11}$ of the hyperfine state $\left|1,0\right\rangle $ can be
freely tuned by an external magnetic field across a Feshbach resonance
\citep{Cabrera2018}. In contrast, the intra-species scattering length
$a_{22}\simeq34a_{0}$ of the hyperfine state $\left|1,-1\right\rangle $
and the inter-species scattering length $a_{12}\simeq-53a_{0}$, where
$a_{0}$ is the Bohr radius, essentially do not change with the magnetic
field. The situation of equal intra-species scattering lengths, considered
in Fig. \ref{fig4_n2n1_a12dep} with a ratio $a_{12}/a\simeq-1.56$,
can be experimentally realized by tuning $a_{11}$ to the value $a_{11}\simeq34a_{0}$.
Similarly, the case of unequal intra-species scattering lengths, shown
in Fig. \ref{fig6_n2n1_a12fix} with $a_{12}/a=-2$, may also be achieved
by tuning $a_{11}$ to $a_{11}\simeq21a_{0}$. Our results therefore
might be testable in experiments, if we do not consider the impact
of the short lifetime of quantum droplets due to the three-body atom
loss \citep{Cabrera2018,Semeghini2018}.

In the two experiments \citep{Cabrera2018,Semeghini2018}, to minimize
the three-body loss, the scattering length $a_{11}$ is typically
tuned to a value $a_{11}\sim68a_{0}$. This corresponds to the case
of $a_{12}/a\simeq-1.1$ shown in Fig. \ref{fig7_n2n1_a12dep}, if
we exchange the indices for the two bosonic components (i.e., $1\leftrightarrow2$).
Therefore, our calculated optimal density ratio differs from the predicted
optimal density ratio in Eq. (\ref{eq:DensityRatioPetrov}) just by
$2\%$, which is too small to be unambiguously identified in the experiments
\citep{Semeghini2018}.

Nevertheless, the situation may dramatically change if the binary
Bose mixture is subjected to a harmonic trapping potential along one
direction (i.e., the $z$-axis), as investigated in the first Barcelona
experiment \citep{Cabrera2018}. Indeed, as revealed by a recent theoretical
study \citep{Cikojevic2020,Cikojevic2021}, a significant derivation
from the predicted optimal density ratio in Eq. (\ref{eq:DensityRatioPetrov}),
as large as $20\%$ in relative, must be taken, in order to reconcile
the theoretical predictions with the experimental data, for the critical
particle number of quantum droplets. In our bosonic pairing theory,
it is possible to take into account the external harmonic trapping
potential, through a numerical modeling of the inhomogeneous pairing
parameter. We would like to leave this heavy numerical calculation
to a future study.

\section{Conclusions}

In summary, using a microscopic pairing theory, we have investigated
the stability window of the droplet state of a binary Bose mixture,
away from the threshold of mean-field collapse. We have shown that,
in the case of unequal intra-species scattering lengths ($a_{11}\neq a_{22}$),
the optimal density ratio may not follow the density locking rule
predicted by Petrov \citep{Petrov2015}, where $n_{2}/n_{1}=\sqrt{a_{11}/a_{22}}$,
as understood and adopted in most of the previous modelings of ultradilulte
quantum droplets. The violation of the density locking rule may also
be seen in the parameter window of the allowed density ratio for the
droplet state, whose size becomes increasingly large with increasing
in the magnitude of the inter-species scattering length $|a_{12}|$.
These observations strongly indicate that the commonly used single-mode
description of quantum droplets, in terms of the extended Gross-Pitaevskii
equation, should be revised, particularly when the inter-species scattering
length $a_{12}$ deviates away from the mean-field collapse threshold.

Our predictions might be tested in future experiments, although a
quantitative confirmation is difficult to obtain due to the potential
severe three-body atom loss, which always happens at the large total
density of quantum droplets. Alternatively, our predictions could
be examined in a systematic investigation with quantum Monte Carlo
simulations \citep{Cikojevic2018,Cikojevic2019,Cikojevic2020,Cikojevic2021},
in which the density ratio of the binary mixture can be changed at
will, so the allowed density ratio of quantum droplets can be directly
determined. 
\begin{acknowledgments}
This research was supported by the Australian Research Council's (ARC)
Discovery Program, Grants Nos. DP240101590 (H.H.), and FT230100229
(J.W.). H.P. acknowledges support from the U.S. NSF (Grant No. PHY-2207283)
and the Welch Foundation (Grant No. C-1669). 
\end{acknowledgments}

\appendix

\section{Stability analysis of the pairing solution}

\begin{figure}
\begin{centering}
\includegraphics[width=0.5\textwidth]{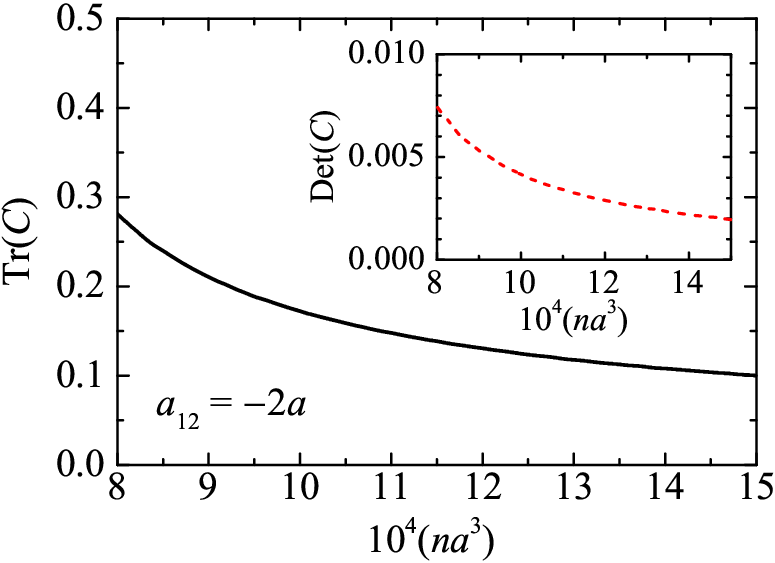}
\par\end{centering}
\caption{\label{fig8_matrixC} The trace (main figure) and the determinant
(inset) of the curvature matrix $\mathcal{C}$, in arbitrary units,
as a function of the total density, at the optimal density ratio $n_{2}/n_{1}=1$.
Here, we consider equal intra-species scattering lengths $a_{11}=a_{22}=a$
and take an inter-species scattering length $a_{12}=-2a$. In this
case, a quantum droplet in free space (with pressure $P=0$) is realized
at $na^{3}\simeq8.9\times10^{-4}$, see, i.e., the black circle in
Fig. \ref{fig3_n2n1_a12fix}.}
\end{figure}

To analyze the stability conditions, it is necessary to calculate
the (symmetric) curvature Hessian matrix, 
\begin{equation}
\mathcal{C}=\left[\begin{array}{cc}
\partial n_{1}/\partial\mu_{1} & \partial n_{1}/\partial\mu_{2}\\
\partial n_{2}/\partial\mu_{1} & \partial n_{2}/\partial\mu_{2}
\end{array}\right].
\end{equation}
A positively defined curvature matrix $\mathcal{C}$ is sufficient
to ensure the mechanical stability (i.e., $\partial n/\partial\mu>0$)
and the spinodal stability (i.e., $\partial\delta n/\partial\delta\mu>0$).
Here, we have introduced the average chemical potential $\mu=(\mu_{1}+\mu_{2})/2$
and the chemical potential difference $\delta\mu=(\mu_{1}-\mu_{2})/2$,
and similarly the total density $n=n_{1}+n_{2}$ and the density difference
$\delta n=n_{1}-n_{2}$. A sufficient and necessary condition of a
positively defined matrix is that its trace and determinant must be
both positive.

In Fig. \ref{fig8_matrixC}, we show the trace and the determinant
of the curvature matrix $\mathcal{C}$, as a function of the total
density at $a_{12}=-2a$ with equal intra-species scattering lengths
$a_{11}=a_{22}=a$. We consider an optimal density ratio $n_{2}/n_{1}=1$.
Both the trace and determinant are positive, confirming that the curvature
matrix is positively defined, in the density range that we are interested
in.

\section{The threshold density ratios from the approximate LHY theory}

\begin{figure}[t]
\begin{centering}
\includegraphics[width=0.5\textwidth]{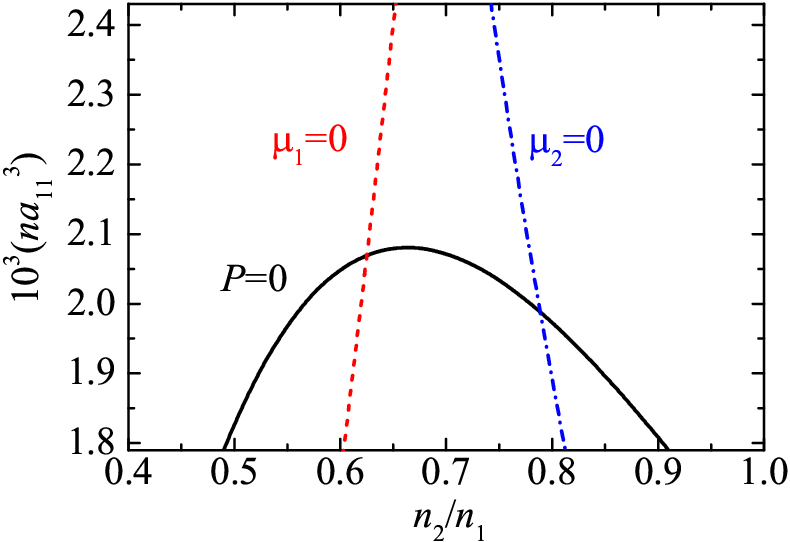}
\par\end{centering}
\caption{\label{fig9_n2n1_LHY} The zero pressure line (the solid line) and
the zero chemical potential lines (the red dashed line and the blue
dot-dashed line), in the plane of the density ratio ($n_{2}/n_{1}$)
and the total density ($na_{11}^{3}$), in the case of unequal intra-species
scattering lengths $a_{11}/a_{22}\simeq0.577$, as predicted by the
approximate LHY theory. As in Fig. \ref{fig6_n2n1_a12fix}, we set
an inter-species scattering length $a_{12}=-2a$, where $a=\sqrt{a_{11}a_{22}}$.
This figure is to be contrasted with Fig. \ref{fig6_n2n1_a12fix},
with the different scale for the total density in mind.}
\end{figure}

Here, we briefly discuss how to calculate the threshold density ratios,
by using the approximate LHY theory as given in Ref. \citep{Zin2021}.
Let us list Eq. (24), Eq. (25) and Eq. (26) of Ref. \citep{Zin2021}
for the pressure and chemical potentials as follows,

\begin{eqnarray}
\frac{P}{e_{0}} & = & -\delta b+\frac{1}{2}\frac{\left(\delta\xi\right)^{2}}{\xi}+\eta\left(\frac{1}{\sqrt{\xi s}}+\sqrt{\xi s}\right)^{5/2},\label{eq:AppendixB_P}\\
\frac{\mu_{1}n_{1}}{e_{0}} & = & -\delta b-\frac{\delta\xi}{\xi}+\frac{5\eta}{3}\left(\frac{1}{\sqrt{\xi s}}+\sqrt{\xi s}\right)^{3/2}\frac{1}{\sqrt{\xi s}},\label{eq:AppendixB_mu1}\\
\frac{\mu_{2}n_{2}}{e_{0}} & = & -\delta b+\frac{\delta\xi}{\xi}+\frac{5\eta}{3}\left(\frac{1}{\sqrt{\xi s}}+\sqrt{\xi s}\right)^{3/2}\sqrt{\xi s},\label{eq:AppendixB_mu2}
\end{eqnarray}
where 
\begin{eqnarray}
e_{0} & \equiv & \frac{4\pi\hbar^{2}}{m}\sqrt{a_{11}a_{22}}n_{1}n_{2},\\
\eta & \equiv & \frac{32}{5\sqrt{\pi}}\left(n_{1}a_{11}^{3}n_{2}a_{22}^{3}\right)^{1/4},\\
s & \equiv & \sqrt{\frac{a_{22}}{a_{11}}},\\
\xi & \equiv & \frac{n_{2}}{n_{1}}\sqrt{\frac{a_{22}}{a_{11}}},
\end{eqnarray}
$\delta\xi\equiv\xi-1$, and $\delta b\equiv-1-a_{12}/\sqrt{a_{11}a_{22}}>0$.
Therefore, the two threshold density ratios can be found by setting
(1) $P=0$ and $\mu_{1}=0$ and (2) $P=0$ and $\mu_{2}=0$.

Let us first consider the lower threshold density ratio determined
by $P=0$ and $\mu_{1}=0$. In this case, we eliminate $\eta$ from
Eq. (\ref{eq:AppendixB_P}) and Eq. (\ref{eq:AppendixB_mu1}) and
obtain, 
\begin{equation}
\delta b=\frac{5\left(\delta\xi\right)^{2}/\left(6\xi\right)+\left(\delta\xi/\xi\right)\left(1+\xi s\right)}{2/3-\xi s}.\label{eq:db1}
\end{equation}
For a fixed ratio $s=\sqrt{a_{22}/a_{11}}$, the left-hand-side of
the above equation is related to the ratio $a_{12}/a$, while the
right-hand-side of the equation is a function of $n_{2}/n_{1}$. Therefore,
for a given ratio $a_{12}/a$, we can uniquely determine the lower
threshold density ratio. Similarly, we have for $P=0$ and $\mu_{2}=0$,
\begin{equation}
\delta b=\frac{5\left(\delta\xi\right)^{2}/\left(6\xi\right)-\delta\xi\left[1+1/\left(\xi s\right)\right]}{2/3-1/\left(\xi s\right)},\label{eq:db2}
\end{equation}
which enables us to determine the upper threshold density ratio.

We have calculated the two threshold density ratios, by solving Eq.
(\ref{eq:db1}) and Eq. (\ref{eq:db2}). The results are shown in
Fig. \ref{fig4_n2n1_a12dep} and Fig. \ref{fig7_n2n1_a12dep} by lines.
We find an excellent agreement between the predictions from the approximate
LHY theory (lines) and the bosonic pairing theory (symbols).

Moreover, in Fig. \ref{fig9_n2n1_LHY} we show the lines for zero
pressure and zero chemical potentials in the plane of the density
ratio ($n_{2}/n_{1}$) and the total density ($na_{11}^{3}$), calculated
with the approximate LHY theory using the same parameters as in Fig.
\ref{fig6_n2n1_a12fix}. We find the essentially same line-shapes
in both figures. The only difference is that the scale for the total
density (i.e., the $y$-axis) is markedly different. In Fig. \ref{fig9_n2n1_LHY},
we need to scale down the total density by a factor of about $5.78$,
in order to match the lines in Fig. \ref{fig6_n2n1_a12fix}.

\end{document}